# Quotient Complexity of Ideal Languages

Janusz Brzozowski[1], Galina Jirásková[2], and Baiyu Li[1]

[1] David R. Cheriton School of Computer Science, University of Waterloo,
Waterloo, ON, Canada N2L 3G1
{brzozo@, b5li@student.cs.}uwaterloo.ca
[2] Mathematical Institute, Slovak Academy of Science,
Grešákova 6, 040 01 Košice, Slovakia
{jiraskov@saske.sk}

October 28, 2018

**Abstract.** We study the state complexity of regular operations in the class of ideal languages. A language $L \subseteq \Sigma^*$ is a right (left) ideal if it satisfies $L = L\Sigma^*$ ($L = \Sigma^* L$). It is a two-sided ideal if $L = \Sigma^* L \Sigma^*$, and an all-sided ideal if $L = \Sigma^* \shuffle L$, the shuffle of $\Sigma^*$ with $L$. We prefer the term "quotient complexity" instead of "state complexity", and we use derivatives to calculate upper bounds on quotient complexity, whenever it is convenient. We find tight upper bounds on the quotient complexity of each type of ideal language in terms of the complexity of an arbitrary generator and of its minimal generator, the complexity of the minimal generator, and also on the operations union, intersection, set difference, symmetric difference, concatenation, star and reversal of ideal languages.

## 1 Ideal Languages

We assume that the reader is familiar with basic concepts of regular languages and finite automata, as described in [14, 18], for example, or in many textbooks. For general properties of ideal languages see [11, 16], for example.

If $\Sigma$ is a non-empty finite alphabet, then $\Sigma^+$ is the free semigroup generated by $\Sigma$, and $\Sigma^*$ is the free monoid generated by $\Sigma$, with empty word $\varepsilon$. A word is any element of $\Sigma^*$. The length of a word $w \in \Sigma^*$ is $|w|$, and $|w|_a$ denotes the number of $a$'s in $w$, where $a \in \Sigma$. A language over $\Sigma$ is any subset of $\Sigma^*$. The *left quotient*, or simply *quotient*, of a language $L$ by a word $w$ is the language $w^{-1}L = \{x \in \Sigma^* \mid wx \in L\}$. *Right quotient* is defined similarly: $Lw^{-1} = \{x \in \Sigma^* \mid xw \in L\}$.

If $u, v, w \in \Sigma^*$ and $w = uv$, then $u$ is a *prefix* of $w$ and $v$ is a *suffix* of $w$. If $w = uxv$ for some $u, v, x \in \Sigma^*$, then $x$ is a *factor* of $w$. Note that a prefix or suffix of $w$ is also a factor of $w$. If $w = w_0 a_1 w_1 \cdots a_n w_n$, where $a_1, \ldots, a_n \in \Sigma$, and $w_0, \ldots, w_n \in \Sigma^*$, then $v = a_1 \cdots a_n$ is a *subword*[3] of $w$; note that every factor of $w$ is a subword of $w$.

A language $L$ is *prefix-convex* if $u, w \in L$ with $u$ a prefix of $w$ implies that every word $v$ must also be in $L$ if $u$ is a prefix of $v$ and $v$ is a prefix of $w$. $L$ is *prefix-free* if $w \in L$ implies that no proper prefix of $w$ is in $L$. $L$ is *prefix-closed* if $w \in L$ implies that every prefix of $w$ is also in $L$. In the same way, we define *suffix-convex*, *factor-convex*, and *subword-convex*, and the corresponding free and closed versions.

A language $L \subseteq \Sigma^*$ is a *right ideal* (respectively, *left ideal*, *two-sided ideal*) if it is non-empty and satisfies $L = L\Sigma^*$ (respectively, $L = \Sigma^*L$, $L = \Sigma^*L\Sigma^*$). We also study special two-sided ideals which satisfy

$$L = \Sigma^* \shuffle L = \bigcup_{a_1 \cdots a_n \in L} \Sigma^* a_1 \Sigma^* \cdots \Sigma^* a_n \Sigma^*,$$

where $\shuffle$ is the shuffle operator. We have not found a name for such an ideal in the literature, so we introduce the term *all-sided ideal*. We refer to all four types as *ideal languages* or simply *ideals*. They have the following properties:

- If $L$ is a right ideal, any $K \subseteq \Sigma^*$ such that $L = K\Sigma^*$ is a *generator* of $L$. The *minimal generator* of $L$ is $G = L \setminus (L\Sigma^+)$, and $G$ is prefix-free.
- If $L$ is a left ideal, any $K \subseteq \Sigma^*$ such that $L = \Sigma^*K$ is a *generator* of $L$. The *minimal generator* of $L$ is $G = L \setminus (\Sigma^+L)$, and $G$ is suffix-free.
- If $L$ is a two-sided ideal, any $K \subseteq \Sigma^*$ such that $L = \Sigma^*K\Sigma^*$ is a *generator* of $L$. The *minimal generator* of $L$ is $G = L \setminus (\Sigma^+L\Sigma^* \cup \Sigma^*L\Sigma^+)$, and $G$ is factor-free.
- If $L$ is an all-sided ideal, any $K \subseteq \Sigma^*$ such that $L = \Sigma^* \shuffle K$ is a *generator* of $L$. The *minimal generator* of $L$ is

$$G = L \setminus \{w \in L \mid \text{ a proper subword of } w \text{ is in } L\},$$

and $G$ is subword-free.

An ideal $L$ is *principal* if it is generated by a language $\{w\}$ consisting of a single word $w \in \Sigma^*$. In that case we write $L = w\Sigma^*$ (rather than $L = \{w\}\Sigma^*$), $L = \Sigma^*w$, *etc.*

Our main interest is in ideal languages that are regular. Left and right ideals were studied by Paz and Peleg [13] in 1965 under the names "ultimate definite" and "reverse ultimate definite events". The results in [13] include closure properties, decision procedures, and canonical representations for these languages. All-sided ideals were used by Haines [8] (not under that name) in 1969 in connection with subword-free and subword-closed languages, and by Thierrin [17] in 1973 in connection with subword-convex languages. De Luca and Varricchio [10] showed in 1990 that a language is factor-closed (also called "factorial") if and only if it is the complement of a two-sided ideal. In 2001 Shyr [16] studied right, left, and two-sided ideals and their generators in connection with codes. In 2008 all four types of ideals were considered by Ang and Brzozowski [1, 2] in the framework of languages convex with respect to arbitrary binary relations. Decision problems for various classes of convex languages, including ideals, were addressed in [6]. Complexity issues of NFA to DFA conversion in right, left, and two-sided ideals were studied in 2008 by Bordihn, Holzer, and Kutrib [3], under the names "ultimate definite", "reverse ultimate definite", and "central definite" languages, respectively.

---
[3] 'Subword' is often used to mean 'factor'; here 'subword' means subsequence.

The closure properties of ideals were analized in [1, 2]. Each of the four classes of ideals is closed under intersection, union, concatenation and inverse homomorphism. Also, right (left) ideals are closed under left (right) quotients, and all-sided ideals are closed under both types of quotients. None of the four classes of ideals is closed under complement, star or homomorphism.

The remainder of the paper is structured as follows. Section 2 explains quotient complexity and describes the derivative approach to finding upper bounds on this complexity. The case of *unary* languages, languages over a one-letter alphabet, is handled in Section 3. The complexity of ideals defined by arbitrary generators and by minimal generators is studied in Sections 4 and 5, respectively, while the complexity of minimal generators is examined in Section 6. The complexities of basic operations on ideals are discussed in Section 7, and Section 8 concludes the paper.

## 2 Quotient complexity

Our approach to quotient complexity follows closely that of [5]. Since the state complexity of a language is a property of a language, it is more appropriately defined in language-theoretic terms. The *quotient complexity* of $L$ is the number of distinct quotients of $L$, and is denoted by $\kappa(L)$.

The following set operations are defined on languages: *complement* ($\overline{L} = \Sigma^* \setminus L$), *union* ($K \cup L$), *intersection* ($K \cap L$), *difference* ($K \setminus L$), and *symmetric difference* ($K \oplus L$). A general *boolean operation* with two arguments is denoted $K \circ L$. We also define the *product*, usually called *concatenation* or *catenation*, ($KL = \{w \in \Sigma^* \mid w = uv, u \in K, v \in L\}$), and *star* ($K^* = \bigcup_{i \geq 0} K^i$). The reverse $w^R$ of a word $w \in \Sigma^*$ is defined as follows: $\varepsilon^R = \varepsilon$, and $(wa)^R = aw^R$. The *reverse* of a language $L$ is denoted $L^R$ and is defined as $L^R = \{w^R \mid w \in L\}$.

*Regular languages* over $\Sigma$ are languages that can be obtained from the *basic languages* $\emptyset$, $\{\varepsilon\}$, and $\{\{a\} \mid a \in \Sigma\}$, using a finite number of operations of union, product and star. Such languages are usually denoted by regular expressions. If $E$ is a regular expression, then $\mathcal{L}(E)$ is the language denoted by that expression. For example, $E = (\varepsilon \cup a)^* b$ denotes $L = (\{\varepsilon\} \cup \{a\})^* \{b\}$.

Since regular languages are denoted by regular expressions, a quotient of a regular language by a word can be denoted by the derivative of the language by that word, as described below.

The $\varepsilon$-*function* $L^\varepsilon$ of a regular expression $L$ is defined as follows::

$$a^\varepsilon = \begin{cases} \emptyset, \text{ if } a = \emptyset \text{ or } a \in \Sigma; \\ \varepsilon, \text{ if } a = \varepsilon. \end{cases} \quad (1)$$

$$(\overline{L})^\varepsilon = \widehat{L^\varepsilon}; \quad (K \cup L)^\varepsilon = K^\varepsilon \cup L^\varepsilon; \quad (KL)^\varepsilon = K^\varepsilon \cap L^\varepsilon; \quad (K^*)^\varepsilon = \varepsilon, \quad (2)$$

where $\widehat{L} = \varepsilon \setminus L$. One verifies that

$$\mathcal{L}(L^\varepsilon) = \begin{cases} \emptyset, & \text{if } \varepsilon \notin L; \\ \{\varepsilon\}, & \text{if } \varepsilon \in L. \end{cases} \quad (3)$$

The *derivative by a letter* $a \in \Sigma$ of a regular expression $L$ is denoted $L_a$ and defined by structural induction:

$$b_a = \begin{cases} \emptyset, & \text{if } b \in \{\emptyset, \varepsilon\} \text{ or } b \in \Sigma \text{ and } b \neq a; \\ \varepsilon, & \text{if } b = a. \end{cases} \quad (4)$$

$$(\overline{L})_a = \overline{L_a}; \quad (K \cup L)_a = K_a \cup L_a; \quad (KL)_a = K_a L \cup K^\varepsilon L_a; \quad (K^*)_a = K_a K^*. \quad (5)$$

The *derivative by a word* $w \in \Sigma^*$ of a regular expression $L$ is denoted $L_w$ and is defined by induction on the length of $w$:

$$L_\varepsilon = L; \quad L_w = L_a, \text{ if } w = a \in \Sigma; \quad L_{wa} = (L_w)_a. \quad (6)$$

A derivative $L_w$ is *accepting* if $\varepsilon \in L_w$; otherwise it is *rejecting*.

Derivatives of a regular expression denote quotients of the language defined by the expression [4,5]:

$$\mathcal{L}(L_w) = w^{-1}L, \text{ for all } w \in \Sigma^*. \quad (7)$$

Two regular expressions are *similar* [4] if one can be obtained from the other using the following rules:

$$L \cup L = L, \quad K \cup L = L \cup K, \quad K \cup (L \cup M) = (K \cup L) \cup M, \quad (8)$$

$$L \cup \emptyset = L, \quad \emptyset L = L\emptyset = \emptyset, \quad \varepsilon L = L\varepsilon = L. \quad (9)$$

Every regular expression has a finite number of dissimilar derivatives [4]. Also, we have $\varepsilon \cap \varepsilon = \varepsilon$, and $\varepsilon \cap \emptyset = \emptyset \cap \varepsilon = \emptyset$.

A *(deterministic, finite) automaton* (DFA) is a quintuple $\mathcal{D} = (Q, \Sigma, \delta, q_0, F)$, where $Q$ is a finite, non-empty set of *states*, $\Sigma$ is a finite, non-empty *alphabet*, $\delta : Q \times \Sigma \to Q$ is the *transition function*, $q_0 \in Q$ is the *initial state*, and $F \subseteq Q$ is the set of *final states*.

The *quotient automaton* of a regular language $L$ is $\mathcal{D} = (Q, \Sigma, \delta, q_0, F)$, where $Q = \{w^{-1}L \mid w \in \Sigma^*\}$, $\delta(w^{-1}L, a) = (wa)^{-1}L$, $q_0 = \varepsilon^{-1}L = L$, and $F = \{w^{-1}L \mid (w^{-1}L)^\varepsilon = \varepsilon\}$. A quotient automaton can be conveniently represented by *quotient equations* [4], which we will use in the simpler notation of derivatives:

$$L_w = \bigcup_{a \in \Sigma} a L_{wa} \cup L_w^\varepsilon,$$

where there is one such equation for each distinct quotient $L_w$ of $L$. Evidently, the number of states in the quotient automaton of $L$ is the quotient complexity of $L$.

A *nondeterministic finite automaton* (NFA) is a tuple $\mathcal{N} = (Q, \Sigma, \eta, S, F)$, where $\eta: Q \times \Sigma \to 2^Q$ and $S \subseteq Q$ is the set of start states.

The following are formulas for the derivatives of regular expressions involving basic operations [4, 5]:

**Proposition 1.** *If $K$ and $L$ are regular expressions, then*

$$(\overline{L})_w = \overline{L_w}, \tag{10}$$

$$(K \circ L)_w = K_w \circ L_w, \tag{11}$$

$$(KL)_w = K_w L \cup K^\varepsilon L_w \cup \left( \bigcup_{\substack{w=uv \\ u,v \in \Sigma^+}} K_u^\varepsilon L_v \right), \tag{12}$$

$$(L^*)_w = \left( L_w \cup \bigcup_{\substack{w=uv \\ u,v \in \Sigma^+}} (L^*)_u^\varepsilon L_v \right) L^*. \tag{13}$$

For notational convenience, $(L_w)^\varepsilon$ is denoted by $L_w^\varepsilon$.

Using the formulas from Proposition 1, we study the quotient complexity of languages of the form $f(L)$ or $f(K,L)$, where $K$ and $L$ are regular ideal languages and $f$ is a regular operation. For simplicity, we use the regular expression notation for both expressions and languages, and the derivative notation for both derivatives and quotients. The meaning is clear from the context.

The next result is from [5]:

**Proposition 2.** *If $\kappa(K) = m$, $\kappa(L) = n$, and $K$ and $L$ have $k > 0$ and $l > 0$ accepting quotients, respectively, then*

1. *If $K$ and $L$ have $\varepsilon$ as a quotient, then*
   - $\kappa(K \cup L) \leq mn - 2$.
   - $\kappa(K \cap L) \leq mn - (2m + 2n - 6)$.
   - $\kappa(K \setminus L) \leq mn - (m + 2n - k - 3)$.
   - $\kappa(K \oplus L) \leq mn - 2$.
2. *If $K$ and $L$ have $\Sigma^+$ as a quotient, then*
   - $\kappa(K \cap L) \leq mn - 2$.
   - $\kappa(K \cup L) \leq mn - (2m + 2n - 6)$.
   - $\kappa(K \setminus L) \leq mn - (2m + l - 3)$.
   - $\kappa(K \oplus L) \leq mn - 2$.
3. *If $K$ and $L$ have $\emptyset$ as a quotient, then*
   - $\kappa(K \cap L) \leq mn - (m + n - 2)$.
   - $\kappa(K \setminus L) \leq mn - n + 1$.
4. *If $K$ and $L$ have $\Sigma^*$ as a quotient, then*
   - $\kappa(K \cup L) \leq mn - (m + n - 2)$.
   - $\kappa(K \setminus L) \leq mn - m + 1$.
5. - *If $L$ has $\varepsilon$ as a quotient, then its reverse $L^R$ has $\kappa(L^R) \leq 2^{n-2} + 1$.*

- If $L$ has $\Sigma^+$ as a quotient, then $\kappa(L^R) \leq 2^{n-2}+1$.
- If $L$ has $\emptyset$ as a quotient, then $\kappa(L^R) \leq 2^{n-1}$.
- If $L$ has $\Sigma^*$ as a quotient, then $\kappa(L^R) \leq 2^{n-1}$.
- Moreover, the effect of these quotients on complexity is cumulative. For example, if $L^R$ has both $\emptyset$ and $\Sigma^*$, then $\kappa(L^R) \leq 2^{n-2}$, if $L^R$ has both $\emptyset$ and $\Sigma^+$, then $\kappa(L^R) \leq 2^{n-3}+1$, etc.

## 3 Unary languages

Unary languages have special properties because the product of unary languages is commutative. Let $\Sigma = \{a\}$. If $L$ is a unary right ideal, let $a^i$ be its shortest word. Then $L \supseteq a^i a^*$, and so $L = a^i a^*$, and every unary right ideal is principal. In fact, $L = a^i a^* = a^* a^i = a^* a^i a^* = a^* \shuffle a^i$; hence left, right, two-sided and all-sided ideals coincide.

**Proposition 3.** *Let $K \subseteq a^*$ and $L \subseteq a^*$ be ideals of any type, with $\kappa(K) = m \geq 1$, $\kappa(L) = n \geq 1$. Let $G$ be the minimal generator of $L$. Then*

$$\kappa(G) = n+1.$$
$$\kappa(L) = \kappa(G) - 1.$$
$$\kappa(K \cup L) = min(m,n).$$
$$\kappa(K \cap L) = max(m,n).$$
$$\kappa(K \setminus L) = \begin{cases} n, & \text{if } m < n; \\ 1, & \text{otherwise.} \end{cases}$$
$$\kappa(K \oplus L) = \begin{cases} max(m,n), & \text{if } m \neq n; \\ 1, & \text{otherwise.} \end{cases}$$
$$\kappa(KL) = m+n-1.$$
$$\kappa(L^*) = \begin{cases} 1, & \text{if } n \in \{1,2\}; \\ n, & \text{otherwise.} \end{cases}$$
$$\kappa(L^R) = n.$$

*Proof.* We prove only the result for $L^*$. If $L \subseteq a^*$ is an ideal with $\kappa(L) = n \geq 1$, then $L = a^{n-1}a^*$. If $n = 1$, then $L = a^* = L^*$, and $\kappa(L) = 1$. If $n = 2$, then $L = aa^*$, $L^* = a^*$, and $\kappa(L) = 1$ again. For $n \geq 3$, $L^* = \varepsilon \cup a^{n-1}a^*$, and $\kappa(L) = n$. □

From now on we usually assume that $|\Sigma| \geq 2$.

## 4 Complexity of ideals in terms of generators

If $L$ is any language and $\trianglerighteq$ is a binary relation on $\Sigma^*$, then the closure of $L$ with respect to this relation [1,2] is $L_{\trianglerighteq} = \{u \mid u \trianglerighteq v \text{ for some } v \in L\}$. If $u \trianglerighteq v$ is the relation "$u$ has $v$ as a prefix", then the closure of $L$ is the right ideal generated by $L$, that is $L_{\triangleright} = L\Sigma^*$. Similarly, if we use the relation "$u$ has $v$ as a suffix (respectively, factor or subword)", then the closure is the left (respectively, two-sided or all-sided) ideal generated by $L$, namely $\Sigma^*L$ (respectively, $\Sigma^*L\Sigma^*$ or $\Sigma^* \shuffle L$). We now investigate the complexity of the closures of any language, that is, the complexity of ideals in terms of arbitrary generators.

## 4.1 Right ideals

The derivative of $KL$ in the case where $L = \Sigma^*$ is:

$$(K\Sigma^*)_w = K_w\Sigma^* \cup K^\varepsilon \Sigma^* \cup \bigcup_{\substack{w=uv \\ u,v \in \Sigma^+}} K_u^\varepsilon \Sigma^*. \tag{14}$$

The following result was shown in [19]; we give a short proof using quotients.

**Theorem 1.** *For any non-empty $K \subseteq \Sigma^*$ with $\kappa(K) = n \geq 1$, we have $\kappa(K\Sigma^*) \leq n$, and the bound is tight.*

*Proof.* If $w$ has no prefix in $K$, then $(K\Sigma^*)_w = K_w\Sigma^*$. Since $K$ is non-empty, there can be at most $n-1$ such quotients, for there must be at least one quotient $K_w$ with $w \in K$. However, for every word $w$ with a prefix $x$ in $K$, we have $(K\Sigma^*)_w = (K\Sigma^*)_x = \Sigma^*$. Hence the bound is $n$.

If $n = 1$ and $\Sigma = \{a\}$, then $K = a^*$ meets the bound. If $n = 2$, use $\Sigma = \{a\}$ and $K = aa^*$. For $n \geq 3$, let $\Sigma = \{a,b\}$ and $K = a\Sigma^{n-3}$. □

## 4.2 Left ideals

The derivative of $KL$ in the case where $K = \Sigma^*$ is:

$$(\Sigma^*L)_w = \Sigma^*L \cup L_w \cup \bigcup_{\substack{w=uv \\ u,v \in \Sigma^+}} L_v. \tag{15}$$

The following result was proved in [19], but the proof there uses a different automaton. In fact, our automaton is the automaton used in [19] for the complexity of the star operation. We include our proof here because we use similar automata later on.

**Theorem 2.** *If $L$ is any language with $\kappa(L) = n \geq 1$, then $\kappa(\Sigma^*L) \leq 2^{n-1}$, and the bound is tight if $|\Sigma| \geq 2$.*

*Proof.* One of the $n$ quotients of $L$, namely $L_\varepsilon = L$, always appears in (15). Thus there are at most $2^{n-1}$ subsets of quotients of $L$ to be added to $\Sigma^*L$.

To prove that the bound is tight, use $\Sigma = \{a\}$ and $L = a^*$ for $n = 1$, and $L = a^*a$ for $n = 2$. For $n \geq 3$ consider the language

$$L = (b \cup a(a \cup b)^{n-1})^*a(a \cup b)^{n-2}.$$

The quotient automaton of $L$ for $n = 5$ is shown in Fig. 1. Then

$$K = \Sigma^* L = \{w \mid w \text{ has an } a \text{ in position } (n-1) \text{ from the end}\}.$$

Let $x$ and $y$ be two different words of length $n-1$, and let $u$ be their longest common prefix. Then, for some $v, w \in \Sigma^*$, we have $x = uav$ and $y = ubw$, and $a^{|u|} \in K_x \setminus K_y$. Hence all the quotients of $K$ by words of length $n-1$ are distinct, and $K$ has at least $2^{n-1}$ distinct quotients. In view of the bound, $K$ has exactly $2^{n-1}$ quotients. □

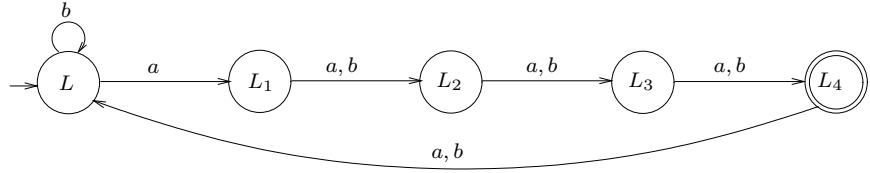

**Fig. 1.** Quotient automaton of $L$ with $\kappa(L) = n = 5$ satisfying $\kappa(\Sigma^* L) = 2^{n-1}$.

In the example of Fig. 1, we have $L_a \neq L$ and $L_b = L$. Since the case $L = L_a = L_b$ leads to $L = \emptyset$ or $L = \Sigma^*$, there are only two more possibilities for quotients by letters, namely: 1) $L_a \neq L_b$, $L_a \neq L$ and $L_b \neq L$, and 2) $L_a = L_b$, $L_a \neq L$. In both of these cases we can improve the bound, as we now show.

**Theorem 3.** *Let $L$ be any language with $\kappa(L) = n \geq 3$. If $L_a \neq L_b$, $L_a \neq L$ and $L_b \neq L$, then $\kappa(\Sigma^* L) \leq 2^{n-1} - 2^{n-3} + 1$, and the bound is tight.*

*Proof.* Note first that this case cannot occur if $n < 3$. Since $L$ always appears in Equation 15, we have at most $2^{n-1}$ subsets of quotients of $L$. Moreover, the quotient $(\Sigma^* L)_{wa}$ always contains $L_a$. Therefore the quotient of $L$ by any word of length greater than zero contains either $L_a$ or $L_b$. Let $S = \{L_1, \ldots, L_{n-1}\}$ be the set of quotients of $L$ other than $L$ itself. There are $2^{n-3} - 1$ non-empty subsets of $S$ containing neither $L_a$ nor $L_b$. These subsets can never appear in the union in Equation 15; hence we have the upper bound.

Consider $n \geq 3$ and the language $L$ defined by the quotient equations:

$$L = aL_1 \cup bL_{n-1},$$
$$L_i = (a \cup b)L_{i+1}, \text{ for } i = 1, 2, \ldots, n-2$$
$$L_{n-1} = aL_1 \cup bL \cup \varepsilon.$$

The quotient automaton of $L$ for $n = 5$ is shown in Fig. 2. Here

$$K = \Sigma^* L = \{w \mid w \text{ ends in } b \text{ or has an } a \text{ in position } (n-1) \text{ from the end}\}.$$

The quotients $K_\varepsilon = K$, $K_{aw}$, where $|w| = n - 2$, and $K_{bva}$, where $|v| = n - 3$, are all distinct: First, we have $b \in K \setminus (K_{aw} \cup K_{bva})$ and $\varepsilon \in K_{aw} \setminus K_{bva}$. Second, consider two different words $x = auaz$ and $y = aubz'$ of the form $aw$; then $a^{|au|} \in K_x \setminus K_y$. Third, consider two different words $x = buaza$ and $y = bubz'a$ of the form $bva$; then $a^{|bu|} \in K_x \setminus K_y$. Thus all the $1 + 2^{n-2} + 2^{n-3}$ quotients are distinct. □

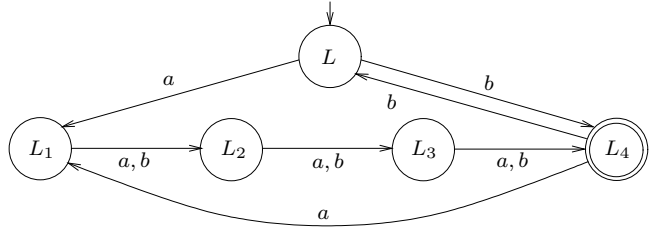

**Fig. 2.** Quotient automaton of $L$ with $\kappa(L) = n = 5$ satisfying $\kappa(\Sigma^*L) = 3 \cdot 2^{n-3} + 1$.

**Theorem 4.** *Let $L$ be any language with $\kappa(L) = n \geq 2$. If $L_a = L_b$ and $L_a \neq L$, then $\kappa(\Sigma^*L) \leq 2^{n-2} + 1$, and the bound is tight.*

*Proof.* Except for $w = \varepsilon$, the quotient $(\Sigma^*L)_w$ always contains $L_a$. Hence the number of possibilities is reduced from $2^{n-1}$ to $2^{n-2} + 1$.

For $n = 2$, let $L = (a \cup b)^*(a \cup b)$; then $\Sigma^*L$ meets the bound. For $n \geq 3$, let $L = \Sigma L'$, where $L'$ is the language in the proof of Theorem 2 (Fig. 1). The quotient automaton of $L$ for $n = 5$ is shown in Fig. 3. Here

$$K = \Sigma^*L' = \{w \mid w \text{ has an } a \text{ in position } (n-2) \text{ from the end}\}.$$

The $1 + 2^{n-2}$ quotients $K_\varepsilon = K$ and $K_{aw}$, where $|w| = n - 2$ are all distinct. Hence the bound is tight. □

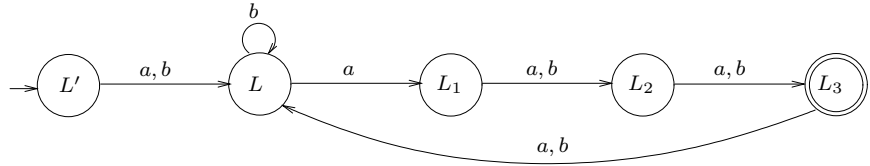

**Fig. 3.** Quotient automaton of $L$ with $\kappa(L) = n = 5$ satisfying $\kappa(\Sigma^*L) = 2^{n-2} + 1$.

### 4.3 Two-sided ideals

Below, it is understood that in $w = uv$ and $v = xy$, we have $u, v, x, y \in \Sigma^+$. If $\varepsilon \notin L$, the derivative of $M = \Sigma^* L \Sigma^*$, is:

$$M_w = \Sigma^*(L\Sigma^*) \cup (L\Sigma^*)_w \cup \bigcup_{w=uv} (L\Sigma^*)_v \tag{16}$$

$$= \Sigma^* L \Sigma^* \cup \left( L_w \Sigma^* \cup \big( \bigcup_{w=uv} L_u^\varepsilon \Sigma^* \big) \right) \cup \bigcup_{w=uv} \left( L_v \Sigma^* \cup \big( \bigcup_{v=xy} L_x^\varepsilon \Sigma^* \big) \right) \tag{17}$$

$$= \left( \Sigma^* L \cup (L_w \cup \bigcup_{w=uv} L_v) \cup \bigcup_{w=uv} \left( L_u^\varepsilon \cup \big( \bigcup_{v=xy} L_x^\varepsilon \big) \right) \right) \Sigma^*. \tag{18}$$

**Theorem 5.** *For every non-empty $L \subseteq \Sigma^*$ with $\varepsilon \notin L$ and $\kappa(L) = n \geq 2$, we have $\kappa(\Sigma^* L \Sigma^*) \leq 2^{n-2} + 1$, and the bound is tight when $|\Sigma| \geq 3$.*

*Proof.* Let $M = \Sigma^* L \Sigma^*$. Since $L$ is always present in the expression $M_w$ above, there are $2^{n-1}$ unions of quotients of $L$ possible. Since $L$ is non-empty, it has at least one accepting quotient. Hence at least $2^{n-2}$ unions contain an accepting quotient of $L$ and the corresponding quotients of $M = \Sigma^* L \Sigma^*$ are $\Sigma^*$. Thus $2^{n-2} + 1$ is an upper bound.

If $n = 2$ and $\Sigma = \{a\}$, then $L = aa^* = a^* aa^*$ meets the bound. If $n = 3$, use $\Sigma = \{a\}$ and $L = aaa^* = a^* aaa^*$. For $n \geq 4$, consider the language $L$ defined by the quotient equations:

$$L = (b \cup c)L \cup aL_1,$$
$$L_i = cL_i \cup (a \cup b)L_{i+1}, \text{ for } i = 1, 2, \ldots, n-3$$
$$L_{n-2} = (a \cup b)L \cup cL_{n-1},$$
$$L_{n-1} = (a \cup b \cup c)L_{n-1} \cup \varepsilon.$$

The quotient automaton of $L$ for $n = 5$ is shown in Fig. 4.

Let $x = uav$ and $y = ubw$ be two different words of length $n - 2$, and let $z = a^{|u|}c$; then $z \in M_x \setminus M_y$. This gives $2^{n-2}$ distinct quotients. Adding $M_{a^{n-2}c} = \Sigma^*$, which is the only quotient of $M$ containing $\varepsilon$, we have the required bound. □

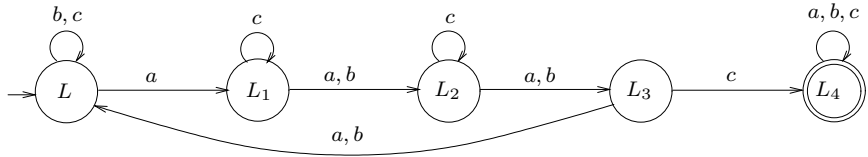

**Fig. 4.** Quotient automaton of $L$ with $\kappa(L) = n = 5$ satisfying $\kappa(\Sigma^* L \Sigma^*) = 2^{n-2} + 1$.

### 4.4 All-sided ideals

The following result was proved by Okhotin [12]. For completeness, we include our short proof.

**Theorem 6.** *For every non-empty $L \subseteq \Sigma^*$ with $\kappa(L) = n \geq 2$, we have $\kappa(\Sigma^* \shuffle L) \leq 2^{n-2} + 1$, and the bound is tight for $n = 2$ if $|\Sigma| \geq 1$ and for $n \geq 3$ if we allow a growing alphabet $\Sigma$ with $|\Sigma| \geq n - 2$.*

*Proof.* Since each all-sided ideal is also two-sided, the bound of $2^{n-2} + 1$ applies.

If $n = 2$, and $\Sigma = \{a\}$, then $L = aa^* = a^*aa^* = \Sigma \shuffle L$ meets the bound. For $n \geq 3$, let $\Sigma = \{a_1, \ldots, a_t\}$, where $t = n-2$, and let $L = \Sigma^*(a_1a_1 \cup \cdots \cup a_ta_t)\Sigma^*$. Then the $n$ distinct quotients of $L$ are $L$, $L_{a_i} = L \cup a_i\Sigma^*$, for $i = 1, \ldots, t$, and $L_{a_1a_1} = \Sigma^*$, since $L_{a_ia_j} = L_{a_j}$ if $i \neq j$ and $L_{a_ia_i} = \Sigma^*$ for all $i$.

Now let $k \geq 0$, $S = \{a_{i_1}, \ldots, a_{i_k}\} \subseteq \Sigma$, where $i_1 < i_2 < \cdots < i_k$, and let $w_S$ be the word $w_S = a_{i_1}a_{i_2}\cdots a_{i_k}$. Thus each letter that is in $S$ appears in $w_S$ exactly once, and the letters are in the order of their subscripts. (For example, for $t = 3$ we have the words $\varepsilon$, $a_1$, $a_2$, $a_3$, $a_1a_2$, $a_1a_3$, $a_2a_3$, and $a_1a_2a_3$.) Also, add the word $a_1a_1$. The quotients of $L$ by these $2^t + 1$ words are all distinct: For two different words $x = a_{i_1}a_{i_2}\cdots a_{i_h}a_iu$ and $y = a_{i_1}a_{i_2}\cdots a_{i_h}a_jv$ with $i < j$, let $z = a_i$; then $z \in L_x \setminus L_y$, since $xz$ contains the letter $a_i$ twice, while all the letters of $yz$ appear only once. Also, $L_{a_1a_1}$ is the only quotient containing $\varepsilon$. Thus $\kappa(\Sigma^* \shuffle L) = 2^{n-2} + 1$, and the bound is tight if we allow a growing alphabet. □

It was also shown in [12] that the bound cannot be reached if an alphabet of only $n - 3$ letters is used.

## 5 Complexity of ideals in terms of minimal generators

Here we consider the following problem: Given a minimal generator $G$ of quotient complexity $n$ for an ideal $L$, what is the complexity of the ideal?

### 5.1 Right ideals

**Theorem 7.** *Let $G$ be the minimal generator of the right ideal $G\Sigma^*$, and let $\kappa(G) = n \geq 3$. Then $\kappa(G\Sigma^*) \leq n$, and the bound is tight if $|\Sigma| \geq 2$.*

*Proof.* The upper bound $n$ follows from [19] or Theorem 1.

For $\Sigma = \{a, b\}$, let $G = a\Sigma^{n-3}$; then $G$ has $n$ quotients and generates the right ideal $L = a\Sigma^{n-3}\Sigma^*$, which also has $n$ quotients. The minimal generator of $L$ is $L \setminus L\Sigma^+ = G\Sigma^* \setminus G\Sigma^*\Sigma^+ = (G \cup G\Sigma^+) \setminus G\Sigma^+ = G$. Hence $G$ is indeed the minimal generator of $G\Sigma^*$, and the bound is tight. □

## 5.2 Left ideals

For $\Sigma = \{a, b\}$, it was stated in [7] that the language $G = a\Sigma^{n-3}$, with $n$ quotients, generates the left ideal $\Sigma^*G$ with $2^{n-2}$ quotients. Since no proof was given that this bound was sufficient, we now provide it.

**Theorem 8.** *Let $G$ be the minimal generator of the left ideal $\Sigma^*G$ and let $\kappa(G) = n \geq 2$. Then $\kappa(\Sigma^*G) \leq 2^{n-2}$, and the bound is tight if $|\Sigma| \geq 2$.*

*Proof.* Let $L = G$ in Equation 15. One of the $n$ quotients of $G$, namely $G_\varepsilon = G$, always appears in the union. Thus there are at most $2^{n-1}$ subsets of quotients of $G$ to be added to $\Sigma^*G$ in (15). Moreover, since $G$ is suffix-free, $G$ has $\emptyset$ as a quotient [9]. Since each union of the $n-1$ quotients other than $G$ that contains $\emptyset$ is equivalent to a union without $\emptyset$, there are at most $2^{n-2}$ quotients of $\Sigma^*G$.

For $n = 2$, let $\Sigma = \{a\}$ and $G = \varepsilon$; then $G$ is the minimal generator of $a^*G = a^*$ and meets the bound.

For $n \geq 3$, let $\Sigma = \{a, b\}$ and $G = a\Sigma^{n-3}$; then $G$ has $n$ quotients, and generates the ideal $L = \Sigma^*G = \{w \mid w \text{ has an } a \text{ in position } (n-2) \text{ from the end}\}$. Thus the $2^{n-2}$ quotients of $L$ by words of length $n-2$ are distinct, and $\kappa(L) = 2^{n-2}$. The minimal generator of $L$ is $L \setminus \Sigma^+L = \Sigma^*G \setminus \Sigma^+\Sigma^*G = (G \cup \Sigma^+G) \setminus \Sigma^+G = G$. Hence $G$ is indeed the minimal generator of $\Sigma^*G$, and the bound is tight. □

## 5.3 Two-sided ideals

**Theorem 9.** *Let $G$ be the minimal generator of the two-sided ideal $\Sigma^*G\Sigma^*$, and let $\kappa(G) = n \geq 3$. Then $\kappa(\Sigma^*G\Sigma^*) \leq 2^{n-3} + 1$, and the bound is tight if $|\Sigma| \geq 2$.*

*Proof.* Let $L = G$ in Equation 18. Since $G_\varepsilon = G$ is always present, there are at most $2^{n-1}$ subsets of quotients of $G$ to add to $G_\varepsilon$. Since $G$ is the minimal generator of $M$, it is factor-free, and hence prefix-free. Thus it has only one accepting quotient, $\varepsilon$, and also has $\emptyset$ as a quotient. So we have at most $2^{n-2}$ subsets, because each subset containing $\emptyset$ is equivalent to another subset without $\emptyset$. Finally, half of those $2^{n-2}$ subsets contain $\Sigma^*$, and hence are equivalent to $\Sigma^*$. This leaves $2^{n-3} + 1$ subsets, and $\kappa(M) \leq 2^{n-3} + 1$.

For $n = 3$, let $\Sigma = \{a\}$ and $G = a$; then $G$ is the minimal generator of $a^*aa^*$ and meets the bound.

For $n \geq 4$, let $\Sigma = \{a, b\}$ and $G = a\Sigma^{n-4}a$; then $G$ has $n$ quotients, and $M = \Sigma^*G\Sigma^*$ has $2^{n-3} + 1$. Then the quotients $M_w$, where $|w| = n - 3$, and $M_{a^{n-2}}$ are all distinct: The only quotient containing $\varepsilon$ is $M_{a^{n-2}}$. If $x = uav$ and $y = ubw$ are two different words of length $n-3$ and $z = a^{|u|}a$, then $z \in M_x \setminus M_y$. The minimal generator of $M$ is

$$M \setminus (\Sigma^+ M \Sigma^* \cup \Sigma^* M \Sigma^+) = \Sigma^* G \Sigma^* \setminus (\Sigma^+ G \Sigma^* \cup \Sigma^* G \Sigma^+) = G.$$

Hence $G$ is indeed the minimal generator of $\Sigma^*G\Sigma^*$ and the bound is tight. □

Theorem 9 shows that the complexity of $\Sigma^* \cdot (G \cdot \Sigma^*)$ is *not* the composition of the complexities of the "double products". By Theorem 7, if $\kappa(G) = n$, then $\kappa(G \cdot \Sigma^*) \le n$. The general bound for the complexity of the product [19] of $K$ and $L$ with $\kappa(K) = m$ and $\kappa(L) = n$ is $m2^n - 2^{n-1}$, which reduces to $2^{n-1}$, when $m = 1$. So the composition of the complexities of the double products yields $2^{n-1}$, whereas the triple product bound is $2^{n-3} + 1$.

### 5.4 All-sided ideals

**Theorem 10.** *Let $G$ be the minimal generator of the all-sided ideal $\Sigma^* \shuffle G$ and let $\kappa(G) = n \ge 4$. Then $k(\Sigma^* \shuffle G) \le 2^{n-3} + 1$, and the bound is tight if we allow a growing alphabet $\Sigma$ with $|\Sigma| \ge n - 3$.*

*Proof.* Since an all-sided ideal is also a two-sided ideal, the bound of $2^{n-3} + 1$ applies.

Let $\Sigma = \{a_1, \ldots, a_t\}$, where $t = n - 3$, and let $G = a_1 a_1 \cup \cdots \cup a_t a_t$. Then $\kappa(G) = n$, the quotients of $G$ being $G$, $G_{a_i} = a_i$, for $i = 1, \ldots, n-3$, $G_{a_1 a_1} = \varepsilon$ and $G_{a_1 a_1 a_1} = \emptyset$.

Now let $L = \Sigma^* \shuffle G = \bigcup_{i=1}^{t} \Sigma^* a_i \Sigma^* a_i \Sigma^*$. Then $L_{a_i} = L \cup \Sigma^* a_i \Sigma^*$, $L_{a_i a_i} = \Sigma^*$ for all $i$, and $L_{a_i a_j} = L_{a_j a_i}$ for all $i, j$. Now let $k \ge 0$, $S = \{a_{i_1}, \ldots, a_{i_k}\} \subseteq \Sigma$, where $i_1 < i_2 < \cdots < i_k$, and let $w_S$ be the word $w_S = a_{i_1} a_{i_2} \cdots a_{i_k}$, as in the proof of Theorem 6, and add the word $a_1 a_1$. The quotients of $L$ by these $2^t + 1$ words are all distinct, as in Theorem 6. Thus $\kappa(L) = 2^{n-3} + 1$.

The minimal generator of $L$ is the set of all words in $L$ that have no proper subwords in $L$. Now $x \in L$ if and only if $x = u a_i v a_i w$ for some $a_i \in \Sigma$, $u, v, w \in \Sigma^*$, and all words of this form are generated by $a_i a_i$. Hence $G$ is indeed the minimal generator of $\Sigma^* \shuffle G$, and the bound is tight if we allow a growing alphabet. □

## 6 Complexity of minimal generators

We now consider the converse problem: Given an ideal $L$ of quotient complexity $n$, what is the quotient complexity of its minimal generator?

**Theorem 11.** *Let $L$ be any right ideal with $\kappa(L) = n \ge 1$. Then the quotient complexity of its minimal generator $G = L \setminus L\Sigma^+$ satisfies $\kappa(G) \le n + 1$, and the bound is tight.*

*Proof.* Since $L = L\Sigma^*$, we have $L\Sigma^+ = L\Sigma$. Now $G_\varepsilon = L \setminus L\Sigma$ and, for $a \in \Sigma$, $x \in \Sigma^*$, $G_{xa} = L_{xa} \setminus (L\Sigma)_{xa}$. By (12), since $\varepsilon \notin L$ because $n > 1$, we have $L^\varepsilon = \emptyset$ and

$$(L\Sigma)_{xa} = L_{xa}\Sigma \cup L^\varepsilon \Sigma_{xa} \cup \left( \bigcup_{\substack{xa=uv \\ u,v \in \Sigma^+}} L_u^\varepsilon \Sigma_v \right) = (L_{xa}\Sigma \cup L_x^\varepsilon \varepsilon), \qquad (19)$$

because the only non-empty quotient of $\Sigma$ by a non-empty word occurs when $v = a$. Thus $G_{xa} = L_{xa} \setminus (L_{xa}\Sigma \cup L_x^\varepsilon \varepsilon)$. We know that, $L$ has only one accepting quotient, which is $\Sigma^*$. If $L_{xa} \neq \Sigma^*$, then $\varepsilon \notin L_{xa}$, and $L_x \neq \Sigma^*$ which implies $L_x^\varepsilon = \emptyset$; thus $G_{xa} = L_{xa} \setminus L_{xa}\Sigma$, and there are $n-1$ such quotients of $G$. If $L_{xa} = \Sigma^*$, then there are two cases:

1. $L_x = \Sigma^*$: we have $\varepsilon \in L_x$ and $G_{xa} = \Sigma^* \setminus (\Sigma^+ \cup \varepsilon) = \emptyset$;
2. $L_x \neq \Sigma^*$: we have $\varepsilon \notin L_x$ and $G_{xa} = \Sigma^* \setminus \Sigma^+ = \varepsilon$. In this case $G_{xa}$ has the form $L_{xa} \setminus L_{xa}\Sigma$, and this has already been counted.

Altogether we have $G_\varepsilon$, $\emptyset$, and $n-1$ other quotients. Hence $\kappa(G) \leq n+1$.

Let $\Sigma = \{a\}$, and let $L = a^{n-1}a^*$, for $n \geq 1$. The $L$ is a right ideal, $\kappa(L) = n$, and the minimal generator is $G = a^{n-1}$ with $\kappa(G) = n+1$. □

*Remark 1.* If $L$ is a left ideal and $u, v \in \Sigma^*$, then $L_v \subseteq L_{uv}$.

*Proof.* If $w \in L_v$, then $vw \in L$. Since $L = \Sigma^*L$, we have $uvw \in L$ for every $u \in \Sigma^*$; so $w \in L_{uv}$ and $L_u \subseteq L_{uv}$. □

**Theorem 12.** *Let $L$ be any left ideal with $\kappa(L) = n \geq 3$. Then the quotient complexity of its minimal generator $G = L \setminus \Sigma^+L$ satisfies $\kappa(G) \leq n(n-1)/2+2$, and the bound is tight if $|\Sigma| \geq 2$.*

*Proof.* Since $L = \Sigma^*L$, we have $\Sigma^+L = \Sigma L$, showing that $G = L \setminus \Sigma L$. Let $L$ be a left ideal with quotients $L_1, L_2, \ldots, L_n$, and let $G$ be the minimal generator of $L$, that is, $G = L \setminus \Sigma L$. If $w = av$ is a nonempty word, then $G_w = L_{av} \setminus L_v$ is a difference of two quotients of $L$. Next, we have $L_v \subseteq L_{av}$, by Remark 1. This means, that if $i \neq j$, then at most one of $L_i \setminus L_j$ and $L_j \setminus L_i$ may be a quotient of $G$. Also, $L_i \setminus L_i = \emptyset$ for all $i$. Hence there are at most $1 + n(n-1)/2 + 1$ quotients of $G$: $G_\epsilon$, at most one quotient for each $i \neq j$, and $\emptyset$.

Next we prove that this bound is tight. Let $n \geq 3$ and let $L$ be the language accepted by the $n$-state DFA of Fig. 5 or denoted by $(b \cup ab)^*a(ab^*)^{n-3}a(a \cup b)^*$. Note that $w \in L$ if and only if $w = xa(ab^*)^{n-3}ay$ for some $x, y \in \Sigma^*$. This follows because every word in $a(ab^*)^{n-3}a$ is accepted from every state of the DFA. Thus $L = \Sigma^*a(ab^*)^{n-3}a\Sigma^*$ is a left ideal.

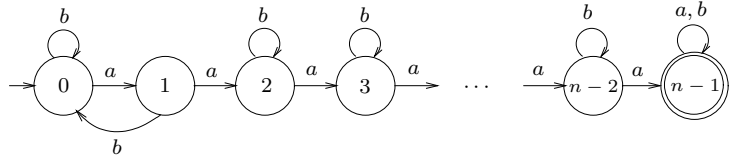

**Fig. 5.** DFA $\mathcal{D}$ for Theorem 12.

Now consider the following $2 + (n-1) + (n-2) + \cdots + 2 + 1 = n(n-1)/2 + 2$ words: $\varepsilon$, $b$, $a(ab)^i a^j$, where $i = 0, 1, \ldots, n-2$ and $j = 0, 1, \ldots, n-2-i$. If $(i,j) \neq (k,l)$, let $x = a(ab)^i a^j$ and $y = a(ab)^k a^l$. We now have several cases:

1. If $i = k$ and $j < l$, take $z = a^{n-2-k-l} = a^{n-2-i-l}$. Then
$$xz = a(ab)^i a^{n-2-i-(l-j)} \notin L,$$
$$yz = a(ab)^k a^l a^{n-2-k-l} = a(ab)^k a^{n-2-k} \in L \setminus \Sigma L.$$

2. If $i = k$ and $j > l$, the argument is symmetric to the first case.
3. If $i < k$, $i + j = k + l$ take $z = a^{n-2-(i+j)}a^k = a^{n-2-(k+l)}a^k$; then $xz \notin L$, $yz \in L \setminus \Sigma L$.
4. If $i < k$, $i + j < k + l$ take $z = a^{n-2-(k+l)}$; then $xz \notin L$ and $yz \in L \setminus \Sigma L$.
5. If $i < k$, $i + j > k + l$ take $z = a^{n-2-(i+j)}$; then $xz \in L \setminus \Sigma L$, $yz \notin L$.

For $\varepsilon$ and $b$ take $z = a^{n-1}$. For $\varepsilon$ and $a(ab)^i a^j$, take $z = a^{n-2-i-j}$, as well as for $b$ and $a(ab)^i a^j$. Hence $G = L \setminus \Sigma L$ has $n(n-1)/2 + 2$ quotients. □

**Theorem 13.** *Let $L$ be any two-sided or all-sided ideal with $\kappa(L) = n \geq 1$. Then the quotient complexity of its minimal generator $G$ satisfies $\kappa(G) \leq n + 1$, and the bound is tight.*

*Proof.* Since every two-sided ideal is a right ideal, the bound of $n + 1$ applies.
Let $\Sigma = \{a\}$, and let $L = \Sigma^* \sqcup a^{n-1}$, for $n \geq 1$. The $L$ is an all-sided ideal, $\kappa(L) = n$, and the minimal generator is $G = a^{n-1}$ with $\kappa(G) = n + 1$. □

## 7 Basic operations on ideals

We now consider the quotient complexity of some basic operations on ideals.

### 7.1 Boolean operations

**Theorem 14.** *If $K$ and $L$ are right ideals (respectively, two-sided ideals, or all-sided ideals) and $\varepsilon \notin K \cup L$, then*

1. $\kappa(K \cap L) \leq mn$,
2. $\kappa(K \cup L) \leq mn - (m + n - 2)$,
3. $\kappa(K \setminus L) \leq mn - (m - 1)$,
4. $\kappa(K \oplus L) \leq mn$.

*If $K$ and $L$ are left ideals, then*

1. $\kappa(K \cap L) \leq mn$,
2. $\kappa(K \cup L) \leq mn$,
3. $\kappa(K \setminus L) \leq mn$,
4. $\kappa(K \oplus L) \leq mn$.

*Furthermore, all these bounds for all ideals are tight.*
*If $\varepsilon \in K \cup L$, then $K \cup L = \Sigma^*$, and $\kappa(K \cup L) = 1$.*

*Proof.* Consider right ideal first.

1. Since $(K \cap L)_w = K_w \cap L_w$, we have $\kappa(K \cap L) \leq mn$. For $m, n \geq 1$ and $\Sigma = \{a, b\}$, the languages $K = (b^*a)^{m-1}\Sigma^*$ and $L = (a^*b)^{n-1}\Sigma^*$ have $\kappa(K) = m$ and $\kappa(L) = n$. The intersection $K \cap L$ consists of all the words that have at least $m - 1$ $a$'s and at least $n - 1$ $b$'s. Since the quotients of $K \cap L$ by the $mn$ words from the set
$$\{a^i b^j \mid 0 \leq i \leq m - 1, 0 \leq j \leq n - 1\}$$
are distinct, the bound is tight.

2. Since $K$ and $L$, both have $\Sigma^*$ as a quotient, by Proposition 2 (4), $\kappa(K \cup L) \leq mn - (m + n - 2)$. For $K$ and $L$ in Part 1 of the proof, the quotients of $K \cup L$ by the $(m - 1)(n - 1) + 1$ words in the set
$$\{a^i b^j \mid 0 \leq i \leq m - 2, 0 \leq j \leq n - 2\} \cup \{a^{m-1}\}$$
are distinct, showing that the bound is tight.

3. Since $K$ and $L$, both have $\Sigma^*$ as a quotient, by Proposition 2 (4), $\kappa(K \setminus L) \leq mn - m + 1$. For $K$ and $L$ in Part 1 of the proof, the quotients of $K \setminus L$ by the $mn - m + 1$ words in the set
$$\{a^i b^j \mid 0 \leq i \leq m - 1, 0 \leq j \leq n - 2\} \cup \{b^{n-1}\}$$
are distinct, showing that the bound is tight.

4. For $K$ and $L$ in Part 1 of the proof, since $(K \oplus L)_w = K_w \oplus L_w$, we have $\kappa(K \oplus L) \leq mn$. Since the quotients of $K \oplus L$ by the $mn$ words from the set
$$\{a^i b^j \mid 0 \leq i \leq m - 1, 0 \leq j \leq n - 1\}$$
are distinct, the bound is tight.

Every all-sided ideal is a two-sided ideal and every two-sided ideal is a right ideal and a left ideal. Thus the upper bound for right ideals also holds for two-sided and all-sided ideals. Also, notice that $K$ and $L$ from Part 1 of the proof are all-sided ideals, for $K = (\Sigma^*a)^{m-1}\Sigma^*$ and $L = (\Sigma^*b)^{n-1}\Sigma^*$. Therefore the theorem holds for two-sided and all-sided ideals also.

For left ideals, the bound $mn$ holds for all four operations, since it holds for regular languages. Since $K$ and $L$ in Part 1 of the proof are left ideals, these bounds are tight for intersection and symmetric difference.

For union let $\Sigma = \{a, b, c, d\}$, and consider $K$ and $L$ defined by the following quotient equations:

$$K = (b \cup c \cup d)K \cup aK_1,$$
$$K_i = (b \cup d)K_i \cup aK_{i+1} \cup cK, \text{ for } i = 1, \ldots, m - 2,$$
$$K_{m-1} = (a \cup b \cup d)K_{m-1} \cup cK \cup \varepsilon.$$

$$L = (a \cup c \cup d)L \cup bL_1,$$
$$L_i = (a \cup c)L_i \cup bL_{i+1} \cup dL, \text{ for } i = 1, \ldots, n - 2,$$
$$L_{n-1} = (a \cup b \cup d)L_{n-1} \cup dL \cup \varepsilon.$$

Consider the quotients of $K \cup L$ by the $mn$ words from the set

$$\{a^i b^j \mid 0 \le i \le m-1, 0 \le j \le n-1\}$$

Clearly $K_{a^i b^j} = K_{a^i}$, and $L_{a^i b^j} = L_{b^j}$. Hence $(K \cup L)_{a^i b^j} = K_{a^i} \cup L_{b^j}$. Therefore all $mn$ quotients of $K \cup L$ are reachable. To prove they are all distinct, notice that $(K \cup L)_{a^i b^j}$ contains the words $a^{m-i-1}$ and $b^{n-j-1}$. If $i, i' < m-1$, $j, j' < n-1$, then this pair of words is only in $(K \cup L)_{a^i b^j}$, but not in any other $(K \cup L)_{a^{i'} b^{j'}}$, if either $i \ne i'$ or $j \ne j'$. If $i = m-1$ or $j = n-1$, then $(K \cup L)_{a^i b^j}$ contains $cb^{n-j-1}$ and $da^{m-i-1}$, and this pair of words is only in this quotient of this type.

For difference of left ideals, we have the bound $mn$. An example that meets the bound is provided by the languages used for the union of left ideals. □

## 7.2 Product

We first state some properties of left ideals.

**Lemma 1.** *If $N = \Sigma^* L$ is a left ideal with $\kappa(N) = r$ and $K$ is any non-empty language with $\kappa(K) = m$, then $\kappa(KN) \le m + r - 1$, and the bound is tight for every $\Sigma$.*

*Proof.* Consider $(KN)_w$. We have $(KN)_\varepsilon = KN = K_\varepsilon N$. If $w \ne \varepsilon$ and there is no factorization $w = uv$ with $u \in \Sigma^*$ and $v \in \Sigma^+$ such that $u \in K$, then $(KN)_w = K_w N$; thus there are at most $m$ such quotients of the form $K_w N$. Note, however, that there is at least one $w \in \Sigma^*$ such that $\varepsilon \in K_w$; otherwise, we would have $K = \emptyset$. For that $w$, we have $K_w N = (K_w \cup \varepsilon) N = K_w N \cup N = K_w N \cup \Sigma^* N = \Sigma^* N = N = N_\varepsilon$. Thus at least one of the $m$ quotients of this type is equal to a quotient of $N$.

Assume now that there is a factorization $w = uv$ with $u \in K$, and let $v'$ be the longest suffix of $w = u'v'$ such that $u' \in K$. By Remark 1,

$$(KN)_w = K_w N \cup K^\varepsilon N_w \cup \bigcup_{\substack{w = uv \\ u, v \in \Sigma^+}} K_u^\varepsilon N_v = K_w N \cup N_{v'}.$$

But $K_w N = K_w \Sigma^* N \subseteq \Sigma^* (\Sigma^* N) = \Sigma^* N = N = N_\varepsilon \subseteq N_{v'}$, and $(KN)_w = N_{v'}$. There are at most $\kappa(N)$ such quotients, but at least one of them has been counted in the first case. Thus there are at most $m + r - 1$ quotients of $KN$.

This bound is met for the one-letter alphabet $\{a\}$. Let $K = a^* \sqcup a^{m-1}$ and $N = a^* \sqcup a^{r-1}$; then $\kappa(K) = m$, $\kappa(N) = r$, and $KN = a^* \sqcup a^{m+r-1}$ has $\kappa(KN) = m + r - 1$. □

**Theorem 15.** *Let $K$ and $L$ be ideals of the same type with $\kappa(K) = m$ and $\kappa(L) = n$. Then the following hold:*

1. *If $K$ and $L$ are right ideals, then $\kappa(KL) \le m + 2^{n-2}$.*
2. *If $K$ and $L$ are left, two-sided, or all-sided ideals, then $\kappa(KL) \le m + n - 1$.*

*Moreover, these bounds are tight.*

*Proof.* In all cases below, let $M = KL$.

1. Suppose that $K = K\Sigma^*$ and $L = L\Sigma^*$ are right ideals. Then $M = KL = K\Sigma^*L\Sigma^* = KN$, where $N = \Sigma^*L\Sigma^*$. Let $\kappa(N) = r$. By Lemma 1, $\kappa(KN) \leq m+r-1$, and our problem reduces to that of finding $r = \kappa(N)$ as a function of $n = \kappa(L)$. By (15) $N_w$ is the union of $\Sigma^*L$ and some quotients of $L$. Since $L$ is always present in the union, we have at most $2^{n-1}$ different unions. Since one of the quotients of $L$ is $\Sigma^*$, and $\Sigma^* \cup L_v = \Sigma^*$, we have at most $2^{n-2}+1$ distinct quotients of $N$. Thus $\kappa(KL) \leq m + 2^{n-2}$.
   To show that the bound is tight, let $\Sigma = \{a,b,c\}$, $K = \Sigma^{m-1}\Sigma^*$, and let $L$ be the right ideal in the proof of Theorem 5. Then $\kappa(K) = m$, $\kappa(L) = n$, $\kappa(\Sigma^*L\Sigma^*) = 2^{n-2}+1$, and $\kappa(KL) = \kappa(\Sigma^{m-1}\Sigma^*L\Sigma^*) = m-1+2^{n-2}+1 = m + 2^{n-2}$.

2. Suppose $K = \Sigma^*K$ and $L = \Sigma^*L$ are left ideals. If $\varepsilon \in K$, then $K = \Sigma^*$, $m = 1$, $KL = L$, and $\kappa(KL) = n = m+n-1$. Otherwise, by Lemma 1, $\kappa(KL) \leq m+n-1$, and this bound is tight.
   Since every all-sided ideal and every two-sided ideal is also a left ideal, the upper bound applies also in these cases. Since our example is an all-sided ideal, the bound is tight in all three cases. □

### 7.3 Star

**Theorem 16.** *If $L$ is an ideal, $\varepsilon \notin L$, and $\kappa(L) = n$, then $\kappa(L^*) \leq n+1$, and this bound is tight for each of the four classes of ideals. If $\varepsilon \in L$ then $\kappa(L^*) = 1$.*

*Proof.* Consider right ideals first. Suppose $\varepsilon \notin L$. If $L = L\Sigma^*$, then $L^* = \varepsilon \cup L\Sigma^*$. Let $M = L^*$. We have $M_\varepsilon = M$, and for $w \in \Sigma^*$,

$$M_w = L_w \Sigma^* \cup \left( \bigcup_{\substack{w=uv \\ u,v \in \Sigma^+}} L_u^\varepsilon (\Sigma^*)_v \right) = \left( L_w \cup \bigcup_{\substack{w=uv \\ u,v \in \Sigma^+}} L_u^\varepsilon \right) \Sigma^*. \qquad (20)$$

Consider $L_w$; if $w = uv$ and $u \in L$, then $L_u = \Sigma^*$, and hence also $L_w = \Sigma^*$. Thus, if $u \in L$, then $L_u = L_w$, and we can use $u$ to define $L_w$. Therefore we can assume that $w$ has no prefix in $L$. In that case, $M_w = L_w\Sigma^*$, and there are at most $n$ such quotients of $M$. So $\kappa(L^*) \leq n+1$.

Since every all-sided ideal is an ideal, and every two-sided ideal is a right ideal, we have an upper bound for these three classes of ideals.

Now let $\Sigma = \{a,b\}$, and $L = (b^*a)^n\Sigma^* = (\Sigma^*a)^n\Sigma^*$. Then the quotients $L_\varepsilon, L_a, \ldots, L_{a^n}$ are distinct, and $\kappa(L) = n+1$. Thus there is an all-sided ideal $L$ such that $\kappa(L^*) = n+1$.

If $L = \Sigma^*L$ is a left ideal, and $\varepsilon \notin L$, then $(L^*)_\varepsilon = \varepsilon \cup LL^*$, and $(L^*)_w$ is given by Equation 13, if $w \in \Sigma^+$. By Lemma 1, $(L^*)_w = L_w L^*$. Hence there are at most $n+1$ quotients of $L^*$. The bound is met for $L = \Sigma^*a^{n-1}$.

Finally, if $\varepsilon \in L$, then $L = \Sigma^*$. □

### 7.4 Reversal

To deal with reversal, we use the well-known subset construction. We start with the quotient automaton $\mathcal{D}$ of $L$, a DFA. We reverse all the transitions of $\mathcal{D}$ to obtain an NFA $\mathcal{N}^R$ accepting $L^R$; the initial state of $\mathcal{D}$ becomes the accepting state of $\mathcal{N}^R$, and the accepting states of $\mathcal{D}$ become the initial states of $\mathcal{N}^R$. The subset construction is then used to obtain a DFA $\mathcal{D}^R$ accepting $L^R$.

**Theorem 17.** *If $L = L\Sigma^*$ is a right ideal and $\kappa(L) = n$, then $\kappa((L\Sigma^*)^R) \leq 2^{n-1}$ and the bound is tight for $|\Sigma| \geq 1$ if $n \in \{1, 2\}$, and for $|\Sigma| \geq 2$ if $n \geq 3$.*

*Proof.* Since $L$ is a right ideal, it has only one accepting quotient, $q_F = \Sigma^*$. This quotient becomes the initial state of the NFA $\mathcal{N}^R$ for $L^R$. Since $L^R$ is a left ideal, we can add a loop for every letter of $\Sigma$ from $q_F$ to $q_F$ in $\mathcal{N}^R$. Therefore $q_F$ appears in every subset of states of $\mathcal{N}^R$ reachable from $q_F$. Hence there are at most $2^{n-1}$ subsets of states of $\mathcal{N}^R$ as states of $\mathcal{D}^R$.

The bound is tight for $\Sigma = \{a\}$ with $n = 1$ for the language $L = a^*$, and with $n = 2$ for $L = aa^*$.

For $n \geq 3$, let $\Sigma = \{a, b\}$ and consider the right ideal $L = (\Sigma^{n-2}b)^*\Sigma^{n-2}a\Sigma^*$ (see Fig. 6 for $L$ with $n = 5$); then $\kappa(L) = n$. Consider the NFA $\mathcal{N}$ obtained by reversing the DFA of $L$. If a word $w$ has length at least $2n - 2$, then it can be accepted and have a $b$ in position $n - 1$ from the end. However, if $|w| \leq 2n - 3$, then $w$ is accepted by $\mathcal{N}$ if and only if $w$ has an $a$ in position $n-1$ from the end. Now, if $x, y \in \Sigma^{n-1}$ and $x = uav, y = ubw$, then $|uava^{|u|}| \leq n-1+n-2 = 2n-3$, since $|u| \leq n-2$. Similarly, $|ubva^{|u|}| \leq 2n-3$. Hence $a^{|u|} \in (L^R)_x \setminus (L^R)_y$. Hence all the quotients of $L^R$ by the $2^{n-1}$ words of length $n - 1$ are distinct. □

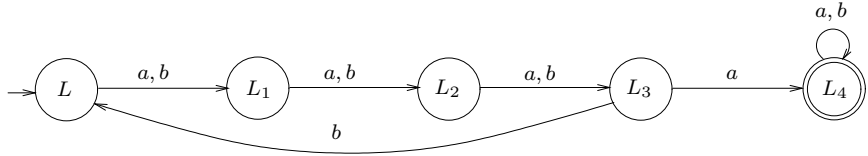

**Fig. 6.** Quotient automaton of $L$ with $\kappa(L) = n = 5$ satisfying $\kappa(L^R) = 2^{n-1}$.

**Theorem 18.** *Let $L$ be a left ideal over an alphabet $\Sigma$ and let $\kappa(L) = n \geq 2$. Then $\kappa((L)^R) \leq 2^{n-1} + 1$ and the bound is tight if $|\Sigma| \geq 3$.*

*Proof.* The quotient $L_\varepsilon = L$, which is the initial state of the quotient automaton $\mathcal{D}$ of $L$, is the only accepting state in the NFA $\mathcal{N}^R$ of $L^R$. In the subset construction, $L$ appears in $2^{n-1}$ subsets. All these subsets are accepting states of $\mathcal{D}^R$, and all accept $\Sigma^*$, since $L^R$ is a right ideal. Hence $\mathcal{D}^R$ has at most $1 + 2^{n-1}$ states.

If $n = 1$, then $\Sigma^*L = \Sigma^*$ and $\kappa((\Sigma^*L)^R) = 1$, for every alphabet. If $n = 2$, the $2^{n-1} + 1$ bound is met for $\Sigma = \{a\}$ and $L = a^*a$. If $n = 3$, the bound is met for $\Sigma = \{a, b, c\}$ and $L = (a \cup b)^*c(c \cup (a \cup b)b^*(a \cup c))^*$.

If $n = 4$, then the bound is met for $\Sigma = \{a, b, c\}$ and $L$ defined by the following quotient equations:

$$L = (a \cup b)L \cup cL_1,$$
$$L_1 = (a \cup b)L_2 \cup cL_1 \cup \varepsilon,$$
$$L_2 = (a \cup b)L_3 \cup cL_1,$$
$$L_3 = (b \cup c)L_1 \cup aL_3.$$

For $n \geq 5$, let $\mathcal{D} = (\{0, 1, \ldots, n-1\}, \{a, b, c\}, \delta, 0, \{n-1\})$ be the DFA shown in Fig. 7. It was proved by Salomaa, Wood, and Yu [15] that the reverse of DFA $\mathcal{D}' = (\{1, \ldots, n-1\}, \{a, b\}, \delta', 1, \{n-1\})$ with $\delta'$ being $\delta$ restricted to $\{a, b\}$, has $2^{n-1}$ states. It follows that the reverse of DFA $\mathcal{D}$ has $2^{n-1} + 1$ states. Since the language accepted by $\mathcal{D}$ is a left ideal, the theorem holds. □

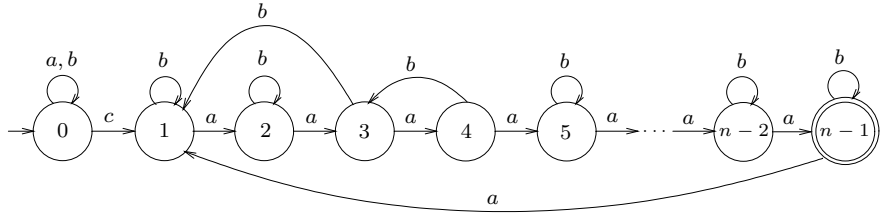

**Fig. 7.** DFA $\mathcal{D}$ for Theorem 18. States $1, 2, \ldots, n-1$ go to state 1 under $c$.

**Theorem 19.** *If $L = \Sigma^*L\Sigma^*$ is a two-sided ideal and $\kappa(L) = n$, $n \geq 2$, then $\kappa((\Sigma^*L\Sigma^*)^R) \leq 2^{n-2} + 1$ and the bound is tight if $|\Sigma| \geq 3$.*

*Proof.* Since $L$ is a right ideal, its quotient automaton $\mathcal{D}$ has exactly one accepting state $q_F$, and this state is not the initial state of $\mathcal{D}$, because $\varepsilon \notin L$. Now $q_F$ is the only initial state of the NFA $\mathcal{N}^R$ accepting $L^R$. Since $L^R$ is a left ideal, we can add a loop for every letter of $\Sigma$ from $q_F$ to $q_F$ in $\mathcal{N}^R$. Therefore $q_F$ appears in every subset of states of $\mathcal{N}^R$ reachable from $q_F$. Hence there are at most $2^{n-1}$ subsets of states of $\mathcal{N}^R$ to consider when using the subset construction.

Since $L$ is a left ideal, the initial state of $\mathcal{D}$ is the only accepting state of $\mathcal{N}^R$ and it appears in $2^{n-2}$ of the subsets of states of $\mathcal{N}^R$. All these subsets are accepting states of $\mathcal{D}^R$, and all accept $\Sigma^*$, since $L^R$ is a right ideal. Hence $\mathcal{D}^R$ has at most $2^{n-2} + 1$ states.

If $n = 1$, then $L = \Sigma^*$ and $\kappa(L^R) = 1$. If $n = 2$, the $2^{n-2} + 1$ bound is met for $\Sigma = \{a\}$ and $L = a^*aa^*$. If $n = 3$, the bound is met for $\Sigma = \{a\}$ and $L = a^*aa^*aa^*$.

For $n \geq 4$, and $\Sigma = \{a, b, c\}$, consider the language $L$ accepted by the $n$-state DFA $\mathcal{D} = (\{0, 1, \ldots, n-1\}, \Sigma, \delta, 0, \{n-1\})$, where $\delta$ is defined in Fig. 8. The language $L$ is a two-sided ideal.

Now construct the NFA for $L^R$. Note that a word $w$ in $(a \cup b)^*c$ of length at most $2n - 4$ is accepted by this NFA if and only if $w$ has an $a$ in the position $n - 1$ from the end and $w$ ends with $c$. We claim that $\{w \in \{a, b\}^* \mid |w| = n - 2\} \cup \{a^{n-2}c\}$ all define distinct quotients. For let $x = uav$ and $y = ubw$ with $|u| \leq n - 3$, and let $z = a^{|u|}c$; then $|xz| = |yz| \leq n - 3 + n - 2 + 1 = 2n - 4$ and $z \in (L^R)_x \setminus (L^R)_y$. Also, $a^{n-2}c$ is in $L^R$ while no $w \in \{a, b\}^*$ is in $L^R$. □

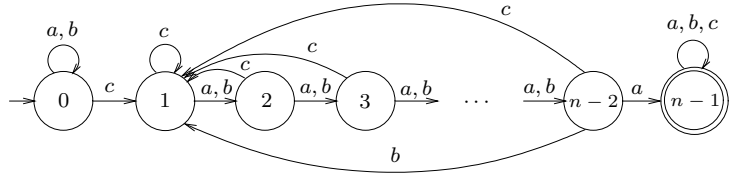

**Fig. 8.** DFA $\mathcal{D}$ for Theorem 19.

**Theorem 20.** *If $L = \Sigma^* \sqcup L$ is an all-sided ideal and $\kappa(L) = n$, $n \geq 2$, then $\kappa((\Sigma^* \sqcup L)^R) \leq 2^{n-2} + 1$ and the bound is tight if we allow a growing alphabet $\Sigma$ with $|\Sigma| \geq 2n - 4$.*

*Proof.* Since an all-sided ideal is also a two-sided ideal, the $2^{n-2} + 1$ bound applies.

If $n = 2$, let $\Sigma = \{a\}$ and $L = a^*aa^*$; then $L = L^R$ and $\kappa(L^R) = 2$. For $n \geq 3$, let $t = n - 2$ and $\Sigma = \{a_1, \ldots, a_t, b_1, \ldots, b_t\}$. Also, let $A = (a_1 \cup \cdots \cup a_t)$, $B = (b_1 \cup \cdots \cup b_t)$, and $B \setminus b_i = (b_1 \cup \cdots \cup b_{i-1} \cup b_{i+1} \cup \cdots \cup b_t)$. Let $L$ be the language defined by the following quotient equations:

$$L = BL \cup \bigcup_{i=1}^{t} a_i L_i,$$
$$L_i = (B \setminus b_i)L_i \cup (A \cup b_i)L_{n-1}, \text{ for } i = 1, \ldots, t,$$
$$L_{n-1} = (A \cup B)L_{n-1} \cup \varepsilon = \Sigma^*.$$

The quotient automaton $L$ for $n = 5$ is shown in Fig. 9. We claim that $L$ is an all-sided ideal; for this, it suffices to show that if $w = uv \in L$ for $u, v \in \Sigma^*$, then $uav \in L$ for every $a \in \Sigma$. If $u = \varepsilon$ and $a \in B$, then $L_a = L$, and if $a = a_i$, then $L_{a_i} = L_i$. However $L_i \supseteq L$; hence all words of the form $\varepsilon av$ are in $L$. If $L_u = \Sigma^*$, then $uav$ is in $L$. Finally suppose that $L_u = L_i$ for some $i$. Since $(L_i)_a$ is either $L_i$ or $\Sigma^*$, it follows that $uav \in L$. Thus $L$ is an all-sided ideal.

Now $L^R = \bigcup_{i=1}^{t}(\Sigma^*(A \cup b_i)(B \setminus b_i)^*a_i B^*)$. Consider the set of $2^{n-2} + 1$ words $\{b_{i_1}b_{i_2} \cdots b_{i_k} \mid 0 \leq k \leq n - 2, 1 \leq i_1 < i_2 < \cdots < i_k\} \cup \{a_1 a_1\}$. If

$x = b_{i_1} \cdots b_{i_l} b_i u$ and $y = b_{i_1} \cdots b_{i_l} b_j v$ with $i < j$, then $a_i \in (L^R)_x \setminus (L^R)_y$. Hence $L^R$ has $2^{n-2} + 1$ quotients. □

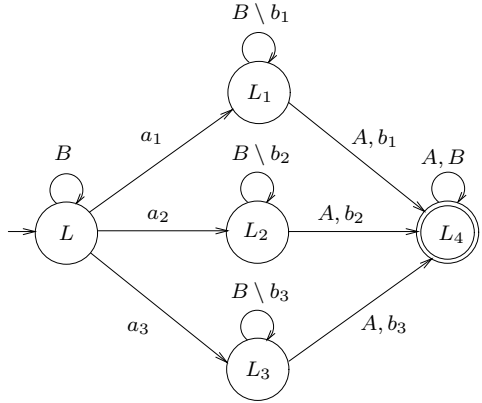

**Fig. 9.** Quotient automaton of all-sided ideal $L$ with $\kappa(L) = n = 5$ satisfying $\kappa(L^R) = 2^{n-2} + 1$.

## 8 Conclusions

Tables 1 and 2 summarize our complexity results. The complexities for regular languages are from [19] (difference and symmetric difference are considered in [5]). In Table 2, $k$ is the number of accepting quotients of $K$ in column $KL$, and the number of accepting quotients of $K$ other than $K$ in column $K^*$.

|  | $K \cup L$ | $K \cap L$ | $K \setminus L$ | $K \oplus L$ |
|---|---|---|---|---|
| unary ideals | $min(m,n)$ | $max(m,n)$ | $n$ | $max(m,n)$ |
| right, 2-sided, all-sided | $mn - (m+n-2)$ | $mn$ | $mn - (m-1)$ | $mn$ |
| left ideals | $mn$ | $mn$ | $mn$ | $mn$ |
| regular languages | $mn$ | $mn$ | $mn$ | $mn$ |

**Table 1.** Bounds on quotient complexity of boolean operations.

**Acknowledgments** This work was supported by the Natural Sciences and Engineering Research Council of Canada under grant no. OGP000871 and by VEGA grant 2/0111/09.

We thank Alexander Okhotin for his help with computations that enabled us to prove Theorem 12.

|           | $f(L)$      | $f(G)$      | $\kappa(G)$           | $KL$            | $K^*$                | $K^R$       |
|-----------|-------------|-------------|-----------------------|-----------------|----------------------|-------------|
| unary     | $n$         | $n-1$       | $n+1$                 | $m+n-1$         | $n-1$                | $n$         |
| right     | $n$         | $n$         | $n+1$                 | $m+2^{n-2}$     | $n+1$                | $2^{n-1}$   |
| 2-sided   | $2^{n-2}+1$ | $2^{n-3}+1$ | $n+1$                 | $m+n-1$         | $n+1$                | $2^{n-2}+1$ |
| all-sided | $2^{n-2}+1$ | $2^{n-3}+1$ | $n+1$                 | $m+n-1$         | $n+1$                | $2^{n-2}+1$ |
| left      | $2^{n-1}$   | $2^{n-2}$   | $\frac{n(n-1)}{2}+2$  | $m+n-1$         | $n+1$                | $2^{n-1}+1$ |
| regular   | $-$         | $-$         |                       | $m2^n-k2^{n-1}$ | $2^{n-1}+2^{n-k-1}$  | $2^n$       |

**Table 2.** Bounds on quotient complexity of generation, product, star and reversal.